\documentstyle[graphicx]{elsart}
\begin{document}
\begin{frontmatter}
\title{Nuclear limits on gravitational waves from elliptically deformed pulsars}
\author{Plamen G. Krastev, Bao-An Li and Aaron Worley}
\address{Department of Physics, Texas A\&M University-Commerce, P.O. Box 3011,\\
Commerce, TX 75429, U.S.A.}
\begin{abstract}
Gravitational radiation is a fundamental prediction of General
Relativity. Elliptically deformed pulsars are among the possible
sources emitting gravitational waves (GWs) with a strain-amplitude
dependent upon the star's quadrupole moment, rotational frequency,
and distance from the detector. We show that the gravitational
wave strain amplitude $h_0$ depends strongly on the equation of
state of neutron-rich stellar matter. Applying an equation of
state with symmetry energy constrained by recent nuclear
laboratory data, we set an upper limit on the strain-amplitude of
GWs produced by elliptically deformed pulsars. Depending on
details of the EOS, for several millisecond pulsars at distances
$0.18kpc$ to $0.35kpc$ from Earth, the {\it maximal} $h_0$ is
found to be in the range of $\sim[0.4-1.5]\times 10^{-24}$. This
prediction serves as the first {\it direct} nuclear constraint on
the gravitational radiation. Its implications are discussed.
\end{abstract}
\end{frontmatter}

\quad{\it Introduction}.-- Gravitational waves are tiny
disturbances in space-time and are a fundamental, although not yet
directly confirmed, prediction of General Relativity. They can be
triggered in cataclysmic events involving (compact) stars and/or
black holes. They could even have been produced during the very
early Universe, well before any stars had been formed, merely as a
consequence of the dynamics and expansion of the Universe. Because
gravity interact extremely weakly with matter, gravitational waves
would carry a genuine picture of their sources and thus provide
undisturbed information that no other messenger can
deliver~\cite{Maggiore:2007}. Gravitational wave astrophysics
would open an entirely new non-electromagnetic window making it
possible to probe physics that is hidden or dark to current
electromagnetic observations~\cite{Flanagan:2005yc}.

\quad (Rapidly) rotating neutron stars could be one of the major
candidates for sources of continuous gravitational waves in the
frequency bandwidth of the LIGO~\cite{Abbott:2004ig} and VIRGO
(e.g. Ref.~\cite{Acernese:2007zzb}) laser interferometric
detectors. It is well known that a rotating object self-bound by
gravity and which is perfectly symmetric about the axis of
rotation does not emit gravitational waves. In order to generate
gravitational radiation over extended period of time, a rotating
neutron star must have some kind of long-living axial
asymmetry~\cite{Jaranowski:1998qm}. Several mechanisms leading to
such an asymmetry have been studied in literature: (1) Since the
neutron star crust is solid, its shape might not be necessarily
symmetric, as it would be for a fluid, with asymmetries supported
by anisotropic stress built up during the crystallization period
of the crust \cite{PPS:1976ApJ}. (2) Additionally, due to its
violent formation (supernova) or due to its environment (accretion
disc), the rotational axis may not coincide with a principal axis
of the moment of inertia of the neutron star which make the star
precess~\cite{ZS:1979PRD}. Even if the star remains perfectly
symmetric about the rotational axis, since it precesses, it emits
gravitational waves \cite{ZS:1979PRD,Z:1980PRD}. (3) Also, the
extreme magnetic fields presented in a neutron star cause magnetic
pressure (Lorenz forces exerted on conducting matter) which can
distort the star if the magnetic axis is not aligned with the axis
of rotation~\cite{BG:1996AA}, which is widely supposed to occur in
order to explain the pulsar phenomenon. Several other mechanisms
exist that can produce gravitational waves from neutron stars. For
instance, accretion of matter on a neutron star can drive it into
a non-axisymmetric configuration and power steady radiation with a
considerable amplitude~\cite{W:1984ApJ}. This mechanism applies to
a certain class of neutron stars, including accreting stars in
binary systems that have been spun up to the first instability
point of the so-called Chadrasekhar-Friedman-Schutz (CFS)
instability~\cite{Schutz:1997}. Also,
Andersson~\cite{Andersson:1997xt} suggested a similar instability
in $r$-modes of (rapidly) rotating relativistic stars. It has been
shown that the effectiveness of these instabilities depends on the
viscosity of stellar matter which in turn is determined by the
star's temperature.

\quad Gravitational wave strain amplitude depends on the degree to
which the neutron star is deformed from axial symmetry which, in
turn, is dependent upon the equation of state (EOS) of
neutron-rich stellar matter. At present time the EOS of matter
under extreme conditions (densities, pressures and isospin
asymmetries) is still rather uncertain and theoretically
controversial. One of the main source of uncertainties in the EOS
of neutron-rich matter is the poorly known density dependence of
the nuclear symmetry energy, $E_{sym}(\rho)$,
e.g.~\cite{Lattimer:2004pg}. On the other hand, heavy-ion
reactions with radioactive beams could provide unique means to
constrain the uncertain density behavior of the nuclear symmetry
energy and thus the EOS of neutron-rich nuclear matter,
e.g.~\cite{Li:1997px,Li:1997rc,Li:2000bj,Li:2002qx,LCK08}.
Applying several nucleonic EOSs, in this letter we calculate the
gravitational wave strain amplitude for selected neutron star
configurations. Particular attention is paid to predictions with
an EOS with symmetry energy constrained by very recent nuclear
laboratory data. These results set an upper limit on the strain
amplitude of gravitational radiation expected from rotating
neutron stars.

\quad The pulsar population is such that most have spin
frequencies that fall below the sensitivity band of current
detectors. In the future, the low-frequency sensitivity of
VIRGO~\cite{Acernese:2005} and Advanced LIGO~\cite{Creighton:2003}
should allow studies of a significantly larger sample of pulsars.
Moreover, LISA  (the Laser Interferometric Space Antenna) is
currently being jointly designed by NASA in the United States and
ESA (the European Space Agency), and will be launched into orbit
by 2013 providing an unprecedented instrument for gravitational
waves search and detection~\cite{Flanagan:2005yc}.

\quad {\it Formalism}.-- In what follows we review briefly the
formalism used to calculate the gravitational wave strain
amplitude. A spinning neutron star is expected to emit GWs if it
is not perfectly symmetric about its rotational axis. As already
mentioned, non-axial  asymmetries can be achieved through several
mechanisms such as elastic deformations of the solid crust or core
or distortion of the whole star by extremely strong misaligned
magnetic fields. Such processes generally result in a triaxial
neutron star configuration~\cite{Abbott:2004ig} which, in the
quadrupole approximation and with rotation and angular momentum
axes aligned, would cause gravitational waves at {\it twice} the
star's rotational frequency~\cite{Abbott:2004ig}. These waves have
characteristic strain amplitude at the Earth's vicinity (assuming
an optimal orientation of the rotation axis with respect to the
observer) of~\cite{HAJS:2007PRL}
\begin{equation}\label{Eq.1}
h_0=\frac{16\pi^2G}{c^4}\frac{\epsilon I_{zz}\nu^2}{r},
\end{equation}
where $\nu$ is the neutron star rotational frequency, $I_{zz}$ its
principal moment of inertia, $\epsilon=(I_{xx}-I_{yy})/I_{zz} $
its equatorial ellipticity, and $r$ its distance to Earth. The
ellipticity is related to the neutron star maximum quadrupole
moment (with $m=2$) via~\cite{Owen:2005PRL}
\begin{equation}\label{Eq.2}
\epsilon = \sqrt{\frac{8\pi}{15}}\frac{\Phi_{22}}{I_{zz}},
\end{equation}
where for {\it slowly} rotating (and static) neutron stars
$\Phi_{22}$ can be written as~\cite{Owen:2005PRL}
\begin{equation}\label{Eq.3}
\Phi_{22,max}=2.4\times
10^{38}g\hspace{1mm}cm^2\left(\frac{\sigma}{10^{-2}}\right)\left(\frac{R}{10km}\right)^{6.26}
\left(\frac{1.4M_{\odot}}{M}\right)^{1.2}
\end{equation}
In the above expression $\sigma$ is the breaking strain of the
neutron star crust which is rather uncertain at present time and
lies in the range $\sigma=[10^{-5}-10^{-2}]$~\cite{HAJS:2007PRL}.
To maximize $h_0$ in this work we take $\sigma=10^{-2}$ which
might be too optimistic. From Eqs.~(\ref{Eq.1}) and (\ref{Eq.2})
it is clear that $h_0$ does not depend on the moment of inertia
$I_{zz}$, and that the total dependence upon the EOS is carried by
the quadrupole moment $\Phi_{22}$. Thus Eq.~(\ref{Eq.1}) can be
rewritten as
\begin{equation}\label{Eq.4}
h_0=\chi\frac{\Phi_{22}\nu^2}{r},
\end{equation}
with $\chi=\sqrt{2045\pi^5/15}G/c^4$. In a recent work we have
calculated the neutron star moment of inertia of both static and
(rapidly) rotating neutron stars. For slowly rotating neutron
stars Lattimer and Schutz~\cite{Lattimer:2005} derived the
following empirical relation
\begin{equation}\label{Eq.5}
I\approx (0.237\pm
0.008)MR^2\left[1+4.2\frac{Mkm}{M_{\odot}R}+90\left(\frac{Mkm}{M_{\odot}R}\right)^4\right]
\end{equation}
This expression is shown to hold for a wide class of equations of
state which do not exhibit considerable softening and for neutron
star models with masses above $1M_{\odot}$~\cite{Lattimer:2005}.
Using Eq.~(\ref{Eq.5}) to calculate the neutron star moment of
inertia and Eq.~(\ref{Eq.3}) the corresponding quadrupole moment,
the ellipticity $\epsilon$ can be readily computed (via
Eq.~(\ref{Eq.2})). Since the global properties of spinning neutron
stars (in particular the moment of inertia) remain approximately
constant for rotating configurations at frequencies up to $\sim
300Hz$~\cite{WKL:2008ApJ}, the above formalism can be readily
employed to estimate the gravitational wave strain amplitude,
provided one knows the exact rotational frequency and distance to
Earth, and that the frequency is relatively low (below $\sim
300Hz$). These estimates are then to be compared with the current
upper limits for the sensitivity of the laser interferometric
observatories (e.g. LIGO).

\begin{figure}[!t]
\centering
\includegraphics[totalheight=4.0in]{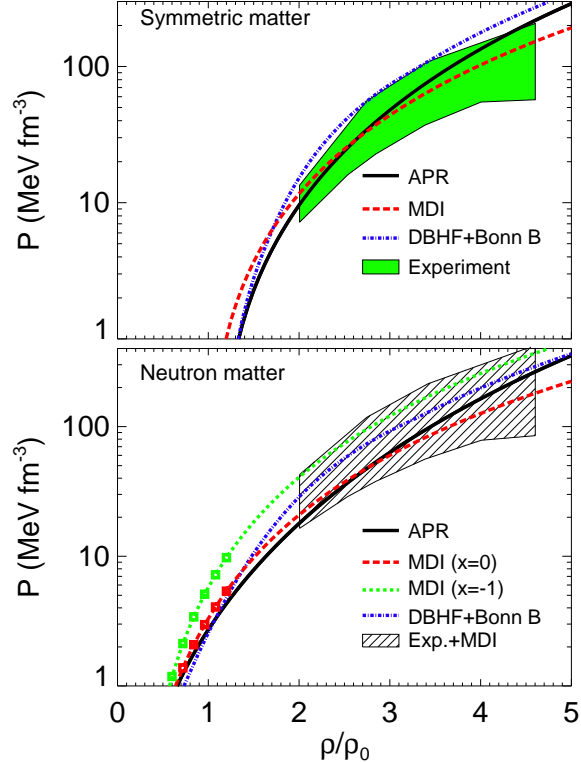}
\vspace{5mm} \caption{(Color online) Pressure as a function of
density for symmetric (upper panel) and pure neutron (lower panel)
matter. The green area in the upper panel is the experimental
constraint on symmetric matter extracted by Danielewicz, Lacey and
Lynch~\cite{DLL:2002Science} from analyzing the collective flow in
relativistic heavy-ion collisions. The corresponding constraint on
the pressure of pure neutron matter, obtained by combining the
flow data and an extrapolation of the symmetry energy functionals
constrained below $1.2\rho_0$ ($\rho_0=0.16 fm^{-3}$) by the
isospin diffusion data, is the shaded black area in the lower
panel. Results are taken partially from
Ref.~\cite{DLL:2002Science}}\label{Fig.1}
\end{figure}

\quad{\it Results and discussion}.-- We calculate the
gravitational wave strain amplitude $h_0$ for several selected
pulsars employing several nucleonic equations of state. We assume
a simple model of stellar matter of nucleons and light leptons
(electrons and muons) in beta-equilibrium. For many astrophysical
studies (as those in this letter), it is more convenient to
express the EOS in terms of the pressure as a function of density
and isospin asymmetry. In Fig.~\ref{Fig.1} we show pressure as a
function of density for two extreme cases: symmetric (upper panel)
and pure neutron matter (lower panel). We pay particular attention
to the EOS calculated with the MDI~\cite{Das:2002fr} (model
dependent) interaction because its symmetry energy has been
constrained in the subsaturation density region by the available
nuclear laboratory data. The EOS of symmetric nuclear matter with
the MDI interaction is constrained by the available data on
collective flow in relativistic heavy-ion collisions. The
parameter $x$ is introduced in the single-particle potential of
the MDI EOS to account for the largely uncertain density
dependence of the nuclear symmetry energy $E_{sym}(\rho)$ as
predicted by various many-body frameworks and models of the
nuclear force. Since it was demonstrated by Li and
Chen~\cite{Li:2005jy} and Li and Steiner~\cite{Li:2005sr} that
only equations of state with $x$ in the range between -1 and 0
have symmetry energy consistent with the isospin-diffusion
laboratory data and measurements of the skin thickness of
$^{208}Pb$, we therefore consider only these two limiting cases in
calculating boundaries of the possible (rotating) neutron star
configurations. It is interesting to note that the symmetry energy
extracted very recently from isoscaling analysis of heavy-ion
reactions is consistent with the MDI calculation of the EOS with
$x=0$~\cite{Shetty:2007}. The MDI EOS has been applied to
constrain the neutron star radius~\cite{Li:2005sr} with a
suggested range compatible with the best estimates from
observations. It has been also used to constrain a possible time
variation of the gravitational constant $G$ \cite{Krastev:2007en}
via the {\it gravitochemical heating} approach developed by Jofre
et al.~\cite{Jofre:2006ug}. More recently we applied the MDI EOS
to constrain the global properties of (rapidly) rotating neutron
stars~\cite{WKL:2008ApJ,KLW2}. In Fig.~\ref{Fig.1} the green area
in the density range of $\rho_0=[2.0-4.6]$ is the experimental
constraint on the pressure $P_0$ of symmetric nuclear matter
extracted by Danielewicz, Lacey and Lynch from analyzing the
collective flow data from relativistic heavy-ion
collisions~\cite{DLL:2002Science}. The pressure of pure neutron
matter $P_{PNM}=P_0+\rho^2dE_{sym}(\rho)/d\rho$ depends on the
density behavior of the nuclear symmetry energy $E_{sym}(\rho)$.
Since the constraints on the symmetry energy from terrestrial
laboratory experiments are only available for densities less than
about $1.2\rho_0$ as indicated by the green and red squares in the
lower panel, which is in contrast to the constraint on the
symmetric EOS that is only available at much higher densities, the
most reliable estimate of the EOS of neutron-rich matter can thus
be obtained by extrapolating the underlying model EOS for
symmetric matter and the symmetry energy in their respective
density ranges to all densities. Shown by the shaded black area in
the lower panel is the resulting best estimate of the pressure of
high density pure neutron matter based on the predictions from the
MDI interaction with $x=0$ and $x=-1$ as the lower and upper
bounds on the symmetry energy and the flow-constrained symmetric
EOS. As one expects and consistent with the estimate in
Ref.~\cite{DLL:2002Science}, the estimated error bars of the high
density pure neutron matter EOS are much wider than the
uncertainty range of the symmetric EOS. For the four interactions
indicated in the figure, their predicted EOSs cannot be
distinguished by the estimated constraint on the high density pure
neutron matter. In addition to the MDI EOS, in Fig.~\ref{Fig.1} we
show results by Akmal et al.~\cite{Akmal:1998cf} with the
$A18+\delta\upsilon+UIX*$ interaction (APR) and recent
Dirac-Brueckner-Hartree-Fock (DBHF)
calculations~\cite{Alonso:2003aq,Krastev:2006ii} with Bonn B
One-Boson-Exchange (OBE) potential (DBHF+Bonn
B)~\cite{Machleidt:1989}. Below the baryon density of
approximately $0.07fm^{-3}$ the equations of state applied here
are supplemented by a crustal EOS, which is more suitable for the
low density regime. Namely, we apply the EOS by Pethick et
al.~\cite{PRL1995} for the inner crust and the one by Haensel and
Pichon~\cite{HP1994} for the outer crust. At the highest densities
we assume a continuous functional for the EOSs employed in this
work. (See~\cite{Krastev:2006ii} for a detailed description of the
extrapolation procedure for the DBHF+Bonn B EOS.) The saturation
properties of the nuclear equations of state applied in this work
are summarized in Table 1.

\begin{table}[!b]
\caption{Saturation properties of the nuclear EOSs (for symmetric
nuclear matter) shown in Fig.~1.}
\begin{center}
\begin{tabular}{lccccc}
EOS &  $\rho_0$ & $E_s$ & $\kappa$ & $m^*(\rho_0)$ & $E_{sym}(\rho_0)$ \\
    &    $(fm^{-3})$ &  $(MeV)$  &  $(MeV)$   & $(MeV/c^2)$ &
    $(MeV)$ \\
\hline\hline
MDI         & 0.160 & -16.08 & 211.00 & 629.08 &  31.62 \\
APR         & 0.160 & -16.00 & 266.00 & 657.25 &  32.60 \\
DBHF+Bonn B & 0.185 & -16.14 & 259.04 & 610.30 &  33.71 \\
\hline
\end{tabular}
\end{center}
\vspace{3mm}{\small The first column identifies the equation of
state. The remaining columns exhibit the following quantities at
the nuclear saturation density: saturation (baryon) density;
energy-per-particle; compression modulus; nucleon effective mass;
symmetry energy.}
\end{table}

\begin{figure}[t!]
\centering
\includegraphics[height=5.5cm]{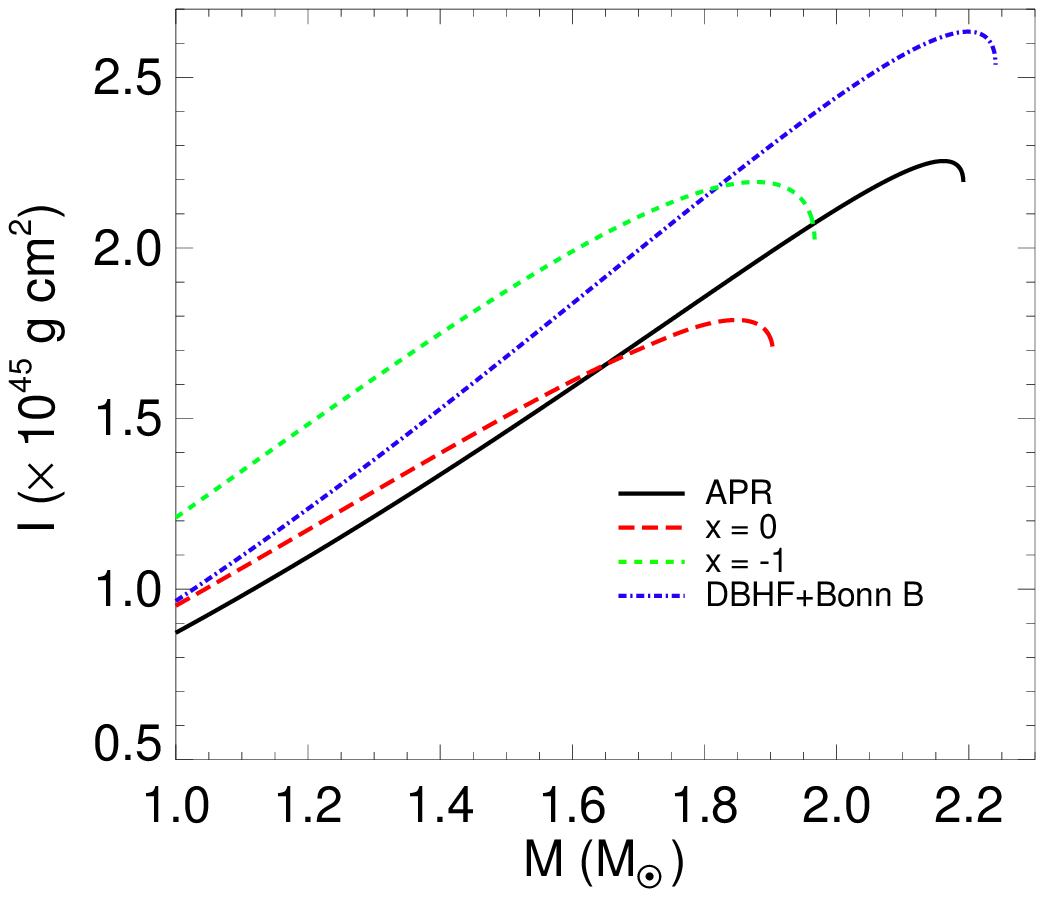}
\includegraphics[height=5.5cm]{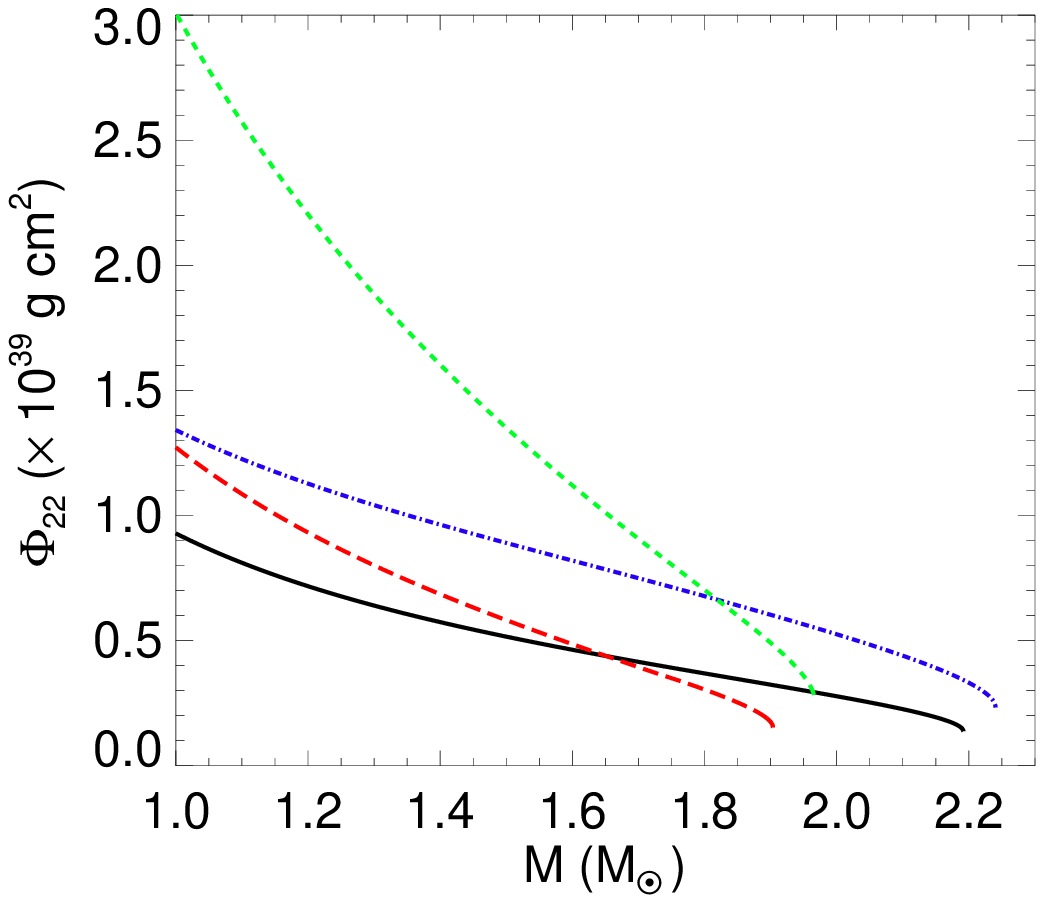}
\caption{(Color online) Neutron star moment of inertia (left
panel; taken from Ref.~\cite{WKL:2008ApJ}) and quadrupole moment
(right panel). $I$ is calculated via Eq.~(\ref{Eq.5}) while
$\Phi_{22}$ via Eq.~(\ref{Eq.3}).} \label{Fig.2}
\end{figure}

\begin{figure}[!b]
\centering
\includegraphics[totalheight=5.5cm]{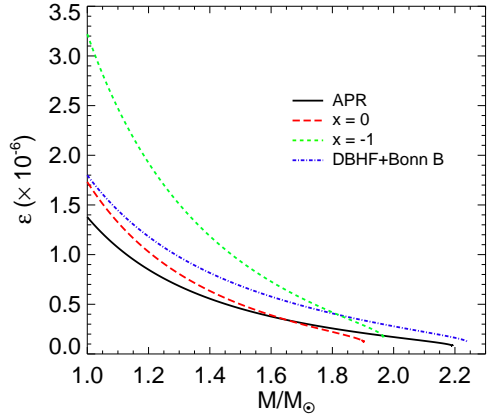}
\vspace{5mm} \caption{(Color online) Ellipticity as a function of
the neutron star mass.} \label{Fig.3}
\end{figure}

\quad Fig.~\ref{Fig.2} displays the neutron star moment of inertia
(left frame) and quadrupole moment (right frame). The moment of
inertia is calculated through Eq.~(\ref{Eq.5}) and the quadrupole
moment through Eq.~(\ref{Eq.3}). Note that these relations are
valid only for {\it slowly} rotating neutron star models. We
notice that $\Phi_{22}$ decreases with increasing stellar mass for
all EOSs considered in this study. The rate of this decrease
depends upon the EOS and is largest for the $x=-1$ EOS. This
behavior is easily understood in terms of the increased central
density with stellar mass -- more massive stars are more compact
and, since the quadrupole moment is a measure of the star's
deformation (see Eq.~(\ref{Eq.2})), they are also less deformed
with respect to less centrally condensed models. Moreover, it is
well known that the mass is mainly determined by the symmetric
part of the EOS while the radius of a neutron star is strongly
affected by the density slope of the symmetry energy. More
quantitatively, an EOS with a stiffer symmetry energy, such as the
$x=-1$ EOS, results in less compact stellar models, and hence more
deformed pulsars. Here we recall specifically that the $x=-1$ EOS
yields neutron star configurations with larger radii than those of
models form the rest of the EOSs considered in this study (e.g.
see Fig.~3 in Ref.~\cite{KLW2}). These results are consistent with
previous findings which suggest that more compact neutron star
models are less altered by rotation, e.g. see
Ref.~\cite{FPI:1984Nature} . Consequently, it is reasonable also
to expect such configurations to be more ``resistant'' to any kind
of deformation. In Fig.~\ref{Fig.3} we show the ellipticity as a
function of the neutron star mass. Since $\epsilon$ is
proportional to the quadrupole moment $\Phi_{22}$ (scaled by the
moment of inertia $I_{zz}$), it decreases with increasing stellar
mass. The results shown in Fig.~\ref{Fig.3} are consistent with
the maximum ellipticity $\epsilon_{max}\approx 2.4\times 10^{-6}$
corresponding to the largest crust ``mountain'' one could expect
on a neutron star~\cite{HAJS:2007PRL,HJA:2006MNRAS}. (The estimate
of $\epsilon_{max}$ in Refs.~\cite{HAJS:2007PRL,HJA:2006MNRAS} has
been obtained assuming breaking strain of the crust
$\sigma=10^{-2}$, as we have assumed in the present paper in
calculating the neutron star quadrupole moment.)

\begin{figure}[!t]
\centering
\includegraphics[totalheight=5.5cm]{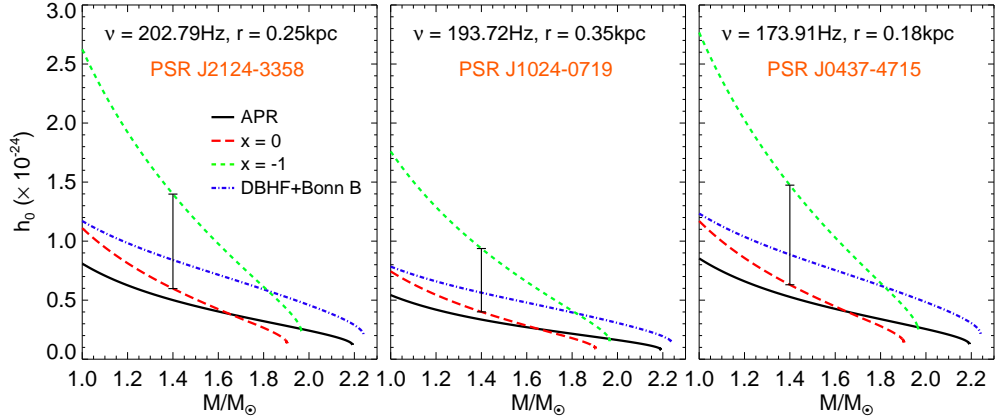}
\vspace{5mm} \caption{(Color online) Gravitational-wave strain
amplitude as a function of the neutron star mass. The error bars
between the $x=0$ and $x=-1$ EOSs provide a limit on the strain
amplitude of the gravitational waves to be expected from these
neutron stars, and show a specific case for stellar models of
$1.4M_{\odot}$.} \label{Fig.4}
\end{figure}

\quad In Fig.~\ref{Fig.4} we display the GW strain amplitude,
$h_0$, as a function of stellar mass. Predictions are shown for
three selected millisecond pulsars, which are relatively close to
Earth ($r<0.4kpc$), and have rotational frequencies below $300Hz$
so that the corresponding momenta of inertia and quadrupole
moments can be computed approximately via Eqs.~(\ref{Eq.5}) and
(\ref{Eq.3}) respectively. The properties of these pulsars (of
interest to this study) are summarized in Table~2.
\begin{table}[!t]
\begin{center}
\caption{Properties of the pulsars considered in this
study.}\vspace{3mm}
\begin{tabular}{lcccc}
\hline
Pulsar        & $\nu(Hz)$ & $M(M_{\odot})$ & $r(kpc)$   & Reference\\
\hline\hline
PSR J2124-3358 & 202.79   & -            &  0.25      &  \cite{Biales:1997ApJ}                   \\
PSR J1024-0719 & 193.72   & -            &  0.35      &  \cite{Biales:1997ApJ}                   \\
PSR J0437-4715 & 173.91   & $1.3\pm 0.2$ &  0.18      &  \cite{Straten:2001Nature,HBO:2006MNRAS} \\
\hline
\end{tabular}
\end{center}
\vspace{3mm} {\small The first column identifies the pulsar. The
remaining columns exhibit the following quantities: rotational
frequency; mass (if known); distance to Earth; corresponding
references. Notice that only the mass of PSR J0437-4715 is known
from orbital dynamics~\cite{Straten:2001Nature,HBO:2006MNRAS} (as
the pulsar has a low-mass white dwarf companion). The masses of
PSRs J2124-3358 and J1024-0719 are presently unknown as they are
both isolated neuron stars~\cite{Biales:1997ApJ}.}
\end{table}
The error bars in Fig.~\ref{Fig.4} between the $x=0$ and $x=-1$
EOSs provide a constraint on the {\it maximal} strain amplitude of
the gravitational waves emitted by the millisecond pulsars
considered here. The specific case shown in the figure is for
neutron star models of $1.4M_{\odot}$. Depending on the exact
rotational frequency, distance to detector, and details of the
EOS, the {\it maximal} $h_0$ is in the range $\sim[0.4-1.5]\times
10^{-24}$. These estimates do not take into account the
uncertainties in the distance measurements. They also should be
regarded as upper limits since the quadrupole moment
(Eq.~(\ref{Eq.3})) has been calculated with $\sigma=10^{-2}$
(where $\sigma$ can go as low as $10^{-5}$). Here we recall that
the mass of PSR J0437-4715 (Fig.~\ref{Fig.4} right panel) is
$1.3\pm 0.2M_{\odot}$ \cite{HBO:2006MNRAS}. (Another mass
constraint, $1.58 \pm 0.18M_{\odot}$, was given previously by van
Straten et al. \cite{Straten:2001Nature}.) The results shown in
Fig.~\ref{Fig.4} suggest that the GW strain amplitude depends on
the EOS of stellar matter, where this dependence is stronger for
lighter neutron star models. In addition, it is also greater for
stellar configurations computed with stiffer EOS. As explained,
such models are less compact and thus less gravitationally bound.
As a result, they could be more easily deformed by rotation or/and
other deformation driving mechanisms and phenomena, and therefore
are expected to emit stronger gravitational radiation
(Eq.~(\ref{Eq.1})).

\begin{figure}[!t]
\centering
\includegraphics[totalheight=6.0cm]{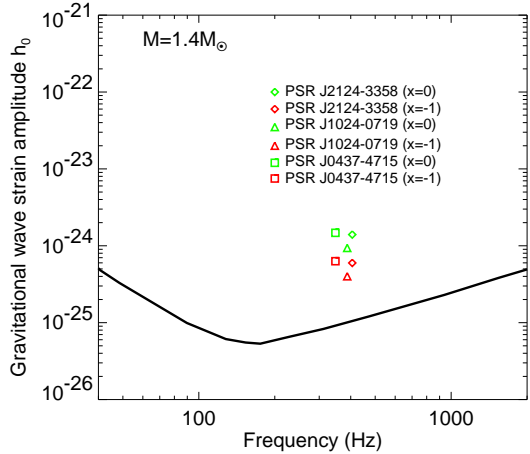}
\vspace{5mm} \caption{(Color online) Gravitational wave strain
amplitude as a function of the gravitational wave frequency. The
characters denote the strain amplitude of the GWs expected to be
emitted from spinning neutron stars ($\nu<300Hz$) with mass
$1.4M_{\odot}$. Solid line denotes the current upper limit of the
LIGO sensitivity. Adapted from Ref.~\cite{Abbott:2004ig}.}
\label{Fig.5}
\end{figure}

\quad In Fig.~\ref{Fig.5} we take another view of the results
shown in Fig.~4. We display the maximal GW strain amplitude as a
function of the GW frequency and compare our predictions with the
best current detection limit of LIGO. The specific case shown is
for neutron star models with mass $1.4M_{\odot}$ computed with the
$x=0$ and $x=-1$ EOSs. Since these EOSs are constrained by the
available nuclear laboratory data they provide a limit on the
possible neutron star configurations and thus gravitational
emission from them. The results shown in Fig.~\ref{Fig.5} would
suggest that presently the gravitational radiation from the three
selected pulsars should be within the detection capabilities of
LIGO. The fact that such a detection has not been made yet
deserves a few comments at this point. First, as we mentioned, in
the present calculation we assume breaking strain of the neutron
star crust $\sigma=10^{-2}$ which might be too optimistic. As
pointed by Haskell et al.~\cite{HAJS:2007PRL}, we are still away
from testing more conservative models and if the true value of
$\sigma$ lies in the low end of its range, we would be still far
away from a direct detection of a gravitational wave signal.
Second, while we have assumed a specific neutron star mass of
$1.4M_{\odot}$, Fig.~\ref{Fig.4} tells us that $h_0$ decreases
with increasing stellar mass, i.e. heavier neutron stars will emit
weaker GWs. Here we recall that from the selected pulsars only the
mass of PSR J0437-4715 is known (within some
accuracy~\cite{Straten:2001Nature,HBO:2006MNRAS}). The masses of
PSR J2124-3358 and PSR J1024-0719 are unknown. Third, in the
present study we assume a very simple model of stellar matter
consisting only beta equilibrated nucleons and light leptons
(electrons and muons). On the other hand, in the core of neutron
stars conditions are such that other more exotic species of
particles could readily abound. Such novel phases of matter would
soften considerably the EOS of stellar medium \cite{BBS:2000PRC}
leading to ultimately more compact and gravitationally tightly
bound objects which could withstand larger deformation forces (and
torques). Lastly, the existence of quark stars, truly exotic
self-bound compact objects, is not excluded from further
considerations and studies. Such stars would be able to resist
huge forces (such as those resulting from extremely rapid rotation
beyond the Kepler, or mass-shedding, frequency) and as a result
retain their axial symmetric shapes effectively dumping the
gravitational radiation (e.g. \cite{Weber:1999a}). At the end, we
recall that Eq.~(\ref{Eq.1}) implies that the best possible
candidates for gravitational radiation (from spinning relativistic
stars) are {\it rapidly} rotating pulsars relatively close to
Earth ($h_0\sim \Phi_{22}\nu^2/r$). Increasing rotational
frequency (and/or decreasing distance to detector, $r$) would
alter the results shown in Figs.~\ref{Fig.4} and \ref{Fig.5} in
favor of a detectable signal by the current observational
facilities (e.g. LIGO). On the other hand, for more realistic and
quantitative calculations, the neutron star quadrupole moment must
be calculated numerically exactly by solving the Einstein field
equations for rapidly rotating neutron stars. (Such calculations
have been reported, for instance, by Laarakkers and
Piosson~\cite{LP:1999ApJ}.) Studies of gravitational waves emitted
from rapidly rotating neutron stars are under way and are left for
a future report.

\quad {\it Summary}.-- We have reported predictions on the upper
limit of the strain amplitude of the gravitational waves to be
expected from elliptically deformed pulsars at frequencies
$<300Hz$. By applying an EOS with symmetry energy constrained by
recent nuclear laboratory data, we obtained an upper limit on the
gravitational-wave signal to be expected from several pulsars.
Depending on details of the EOS, for several millisecond pulsars
$0.18kpc$ to $0.35Kpc$ from Earth, the {\it maximal} $h_0$ is
found to be in the range of $\sim[0.4-1.5]\times 10^{-24}$. This
prediction sets the first direct nuclear constraint on the
gravitational waves from elliptically deformed pulsars.

\quad {\it Acknowledgements}.-- The authors gratefully acknowledge
support from the National Science Foundation under Grant No.
PHY0652548, the Research Corporation under Award No. 7123 and the
Texas Coordinating Board of Higher Education grant No.
003565-0004-2007.

\end{document}